\documentclass[11pt,twoside,a4paper]{article}

\usepackage[american]{babel}
\usepackage{bm}
\usepackage{amsfonts}
\usepackage{graphicx}
\usepackage{amsmath}

\newcommand{\nn}{\nonumber}
\newcommand{\bs}[1]{\boldsymbol{#1}}

\usepackage{color}

\newcommand{\zd}{\delta}
\newcommand{\zD}{\Delta}
\newcommand{\zs}{\sigma}

\newcommand{\ze}{\varepsilon}
\newcommand{\zb}{\beta}
\newcommand{\zg}{\gamma}

\newcommand{\zm}{\mu}

\newcommand{\zr}{\rho}

\newcommand{\zy}{\psi}

\newcommand{\dif}{\; \textrm d }

\newcommand{\dpp}[2]{\frac{  \partial #1 }{  \partial #2   } }
\newcommand{\dpt}[2]{\frac{  \dif #1 }{  \dif #2   } }

\usepackage{a4wide}

\begin{document}

\begin{center}
\Large{\textbf{Stochastic model for quantum spin dynamics in magnetic nanostructures}}\\
	\small{O. Morandi}	\\
\vskip0.5cm
	\textit{\textsf Dipartimento di Matematica e Informatica U. Dini, \\Universit\`a di Firenze,\\ Viale Morgagni 67/A, 50134 \\
		Firenze, Italy.}
\vskip0.5cm
	\textit{omar.morandi@unifi.it}
\end{center}

\begin{center}
\begin{minipage}[h]{0.8\textwidth}
 \small

We develop a numerical model that reproduces the  thermal equilibrium and the spin transfer mechanisms in magnetic nanomaterials. We analyze the coherent two-particle spin exchange interaction and the electron-electron collisions.
Our study is based on a quantum atomistic approach and the particle dynamics is performed by using a Monte Carlo technique.    
 The coherent quantum evolution of the atoms is interrupted by instantaneous collisions with itinerant electrons. The collision processes are associated to the quantum collapse of the local atomic wave function. We show that particle-particle interactions beyond the molecular field approximation can be included in this framework. 

Our model is able to reproduce the thermal equilibrium  and strongly out-of-equilibrium  phenomena such as the ultrafast dynamics of the magnetization in nanomatrials.

\end{minipage}
\end{center}
\vspace{1cm}
\normalsize

\section{Introduction}

The study of the  magnetism dynamics in a nanostructure is a rapidly developing area of research. The technological interest relies on the applications to magnetic data storage and sensing devices. The conventional way to record informations in a magnetic memory is to modify the local magnetization of a magnetic material \cite{Bigot_13}.
Understanding the spin dynamics in magnetic materials is an issue of crucial importance for progress in information processing and recording technology.
The use of a laser field has been proved to lead to a considerable speed up of the digital information recording. The elementary mechanisms that lead to the modification of the magnetic order triggered by laser pulses are at the center of an intense debate \cite{Zhang_00,morandi_PRE_14,Baral_14,Manchon_12, Koopmans_10,Boeglin_10,Battiato_10,Carva_11,Schellekens_13,Wienholdt_13,Bergeard_14}.

The physical properties of a magnetic system arise from the microscopic configuration of the angular momentum and the spin of the atoms.
%
%
It is generally challenging to develop a framework capable of reproducing two important aspects of the nano-material physics: the coherent quantum mechanical evolution of the system and the loss of quantum coherence due to collisions. Usually, the coherent quantum evolution of atoms and electrons is described at the single particle level  by reversible Hamiltonian dynamics. Collisions are a complex many-particle effect and are usually approximated by classical mechanics or by a framework where quantum mechanics plays a minor role. As an example, in the Boltzmann kinetic theory the quantum mechanics is limited to the calculation of scattering parameters such as the effective cross sections. 
Concerning the study of a nanometric systems, it seems natural to rely on a full ab initio many-body description, which is able in principle to describe at the same time the collisional and the purely coherent quantum mechanical evolution of electrons and atoms. However, due to the inherent extensive numerical effort, in many cases the application of ab initio  approaches is  limited  to periodic structures or to small particle systems like molecules  \cite{Krieer_15}. 

Such difficulties have encouraged the development of heuristic approaches. Coarse-grained methods and discrete mesoscopic models are adopted
to assess the properties of materials at atomistic or continuum level. 
The so called atomistic models provide a simple and powerful framework capable to reproduce the microscopic spin dynamics of complex systems \cite{Evans_14,Skubic_14}. 
Atomistic models are based on the classical description of the atomic spin provided by the phenomenological Landau-Lifshitz-Gilbert (LLG) equation.
They have been used to reproduce the ultrafast evolution of the total magnetization in systems excited by laser pulses and the angular momentum  transfer in alloys containing metals and rare earth elements \cite{Radu_11}.
Such models suffer of a serious theoretical limitation that reduces considerably their area of application. Atomistic models for magnetic materials are based  on the classical spin theory. Electron spin is intrinsically a quantum mechanical phenomena, the classical description being valid only in the limit of large spin.

In the present contribution, we propose an atomistic-like approach based on the quantum mechanical framework. A cluster of atoms or, more generally, a nanomaterial is described by a network of quantum spin particles interacting via exchange interaction. We treat also the interaction of such particles with a gas of itinerant electrons. 
The main concern of our approach is to develop a Monte Carlo (MC) stochastic method able to reproduce the twofold aspect of the mesoscopic dynamics, the coherent and the collisional one. 
We observe that both the quantum and the classical collisional dynamics can be modeled by time-continuous stochastic processes. 
The central point of our approach is to assume that every collision among bounded atomic electrons and itinerant ones causes the Heisenberg collapse  of the atomic particle wave function. In fact, the Heisenberg  quantum measurement process is expected to take place every time a quantum system  interacts with a classical apparatus or with an external system containing many degree of freedom.  
It this regards, we treat the delocalized electrons as a classical bath interacting with the open quantum system constituted by the net of interacting atoms.


Our model is based on the following scheme. We consider a set of localized particles each of one is described by a particle wave function $\psi^i$ that evolves according to the reversible Hamiltonian dynamics obtained by a local model of the atom-atom interaction. 
The coherent evolution is interrupted by instantaneous collision events. We assume that the atomic wave function collapses in a well defined  state according to the quantum measurement paradigm. 
%
%
According to our model of collision, we simulate the modification of the atomic state and we estimate the post-collision wave function. 
The collision times and the modification of the particle state induced by the collisions, are evaluated by generating random of numbers that, in the spirit of the MC technique, reproduce the collision processes. 

In Ref. \cite{Morandi_PRB17} this model has been applied to reproduce the ultrafast magnetization dynamics in a Co$_x$Tb$_{1-x}$ alloy . We considered the evolution of the spin and of the  orbital angular momentum of a large number of representative  atoms of the alloy. The simulations showed that our atomistic model is able to treat the spin-orbit interaction, the atomic magnetic anisotropy and the interaction of atoms with phonons mediated by itinerant electrons. Here, we consider a simplified model where only the atomic spin and exchange interaction are included. Our main concern is to provide a comprehensible description of the simulation details. 

We illustrate our method by calculating the value of the mean magnetization of a ferromagnet at the equilibrium and the out-of-equilibrium magnetization dynamics.

\section{Local atom-atom interaction: coherent evolution}\label{sec_NN_int}

We divide the particle dynamics in two parts, collisional and coherent dynamics. The coherent one comprises two phenomenas, $i$) the short range atom-atom interaction treated at the molecular field level and $ii$) the two-body exchange interaction. In this section we discuss the point $i$) and we postpone the discussion concerning the point $ii$) to Sec. \ref{sec_two-b int}.


According to the Heisenberg theory, a ferromagnet can be view as an array of interacting spins. The interaction length is usually very short (contact interaction) and the spin-spin interaction is limited to few neighbors of the atom. We study a nanosystem containing  $N$ atoms. Each atom is described  by the wave function $\psi^i\in H^i$ where $i=1,\ldots,N$, $H^i$ is the atomic Hilbert spin space with dimension $2S^i+1$ and $S^i$ is the total spin of the  atom. The evolution of the atomic wave function  is given by the following Schr\"odinger equation
\begin{align}
 i\hbar \dpp{\zy^i}{t} = \left(\frac{g \zm_B}{\hbar} \mathbf{B}_0  -    \frac{ 1 }{   \hbar^2   }    \sum_{\langle   j\in \textrm{ NA}_i \rangle} \zg_{ij} \mathbf{S}^j \right) \cdot \mathbb{S}^i  \zy^i \;. \label{Sch NN}
\end{align}
The first term in the right-hand side takes into account the Zeeman interaction due to an external magnetic field $ \mathbf{B}_0$, $\mu_B$ is the Bohr magneton and $g$ the Land\'e g-factor. $ \mathbb{S}^i$ denotes the spin matrix of the atom at the position $\mathbf{R}^i$. The spin operator obeys to the commutation rule
\begin{align}
[ \mathbb{S}_i^r,\mathbb{S}_j^s]=i \hbar\; \zd_{r,s}  \ze_{ijk} \; \mathbb{S}_k^r \quad\textrm{ with } i,j,k = x,y,z \; ,
\end{align}
where $\ze$ denotes the antisymmetric Ricci tensor and $\zd$ the Kronecker delta. The low index  refers to the component of the spin operator along the coordinate axes. For a one-half spin particle, $S= \frac{1}{2}$ and $\mathbb{S}= \frac{\hbar}{2} \bs \zs $, where $\bs \zs $ is the vector having Pauli matrices as its components. The second term in the right-hand side  of Eq. \eqref{Sch NN} describes the spin-spin exchange interaction between two neighbor atoms. The strength of the exchange interaction is given by the exchange matrix $\zg_{ij}$ and the symbol NA$_i$ indicates that the sum is taken over the neighbors of the $i-$th atom. The vector $ \mathbf{S}^i  $ denotes the atomic spin.  $ \mathbf{S}^i  $ is   the expectation value of the operator $ \mathbb{S}^i $
\begin{align}
\mathbf{S}^i   \equiv  \langle\zy^i|  \mathbb{S}^i | \zy^i \rangle \;,  \label{expect S}
\end{align}
where we have used the bracket Dirac notation.
Equation \eqref{Sch NN} is very popular and  reproduces with  good approximation the local atomic energy structure in nanosystems and molecules.
Concerning the application to magnetic nanosystems, the local interaction described by Eq. \eqref{Sch NN}  has been applied to reproduce the spin dynamics in quantum and classical models \cite{Evans_14,Zhang_13,Frietsch_13}. 
Equation \eqref{Sch NN} can be easily extended to include other phenomenas as for example the spin-orbit interaction, the presence of the lattice potential, the anisotropy field and of the orbital quenching effect \cite{Morandi_PRB17}.

\section{Collision between localized and itinerant electrons}\label{sec_atom_free_int}

In this section we describe how we include in our model the collisions between the electrons localized around the atomic sites and the itinerant electrons. The itinerant electrons are treated as a gas and are described by a quasi-equilibrium density function. 

It is convenient to describe the collision between atoms and itinerant electrons by using the density matrix formalism \cite{Baral_14,Cywiski_07}. We take  particular care to maintain the coherence between the density matrix approach and the single particle description of the atomic wave function adopted in the previous section. 
We denote by $n_m$ the diagonal components of the single particle atomic density matrix.  This quantities represent the probability that the $m-$th state is occupied. In this section, we assume that the atomic spin is quantized along the $z-$axes. If we assume that the density matrix is diagonal,  the  local spin magnetization orthogonal to the $z-$axes  is zero (collinear case). We have
\begin{align}
S_z = \left\langle \psi |\mathbb{S}_z|\psi \right\rangle = \sum_{m} m |\psi_m|^2=\sum_{m} m n_m\;. \label{magn spin dens}
\end{align}
On order to simplify the notation, in this section we suppress the index related to the atomic position in the lattice. The evolution equation for the occupation probability is
\begin{align}
\left. \dpt{n_m}{t} \right|_{l-i} =
&    W^{m}_+n_{m+1}  +    W^{m}_{-} n_{m-1}     - W^{m}_0   n_m  \equiv \sum_{m'} Q_{m,m'} n_{m'} \label{kolmog_1}\;.
\end{align} 
We have defined  
\begin{align*}
W^{m}_+= 
&     \mathcal{T} w_{-}^m   e^{ \zb\frac{\zD}{2} }           
 \\
W^{m}_-= &    \mathcal{T} w_{+}^{m}          e^{-\zb\frac{\zD}{2} }      \\
W^{m}_0= &   \mathcal{T}  \left(w_{-}^m   e^{-\zb\frac{\zD}{2} }   + w_{+}^{m}    e^{ \zb\frac{\zD}{2} }    \right) \;,
\end{align*}
where $ w_{\pm}^m  = \frac{\zg^2}{\hbar} \left( S(S+1)- m(m \mp 1)\right)$, $S$ is the total atomic spin, $\zg$ is the exchange interaction strength, $\zb=\frac{1}{k_BT} $ where $T$ is the temperature of itinerant electrons and $k_B$ the Boltzmann's constant. The symbol $\zD=  \ze_{m+1} - \ze_{m } +\mu_\downarrow-\mu_\uparrow $ expresses the energy balance during the collision process, $\ze_m$ indicates the energy of the local atomic Hamiltonian for the state $m$ obtained by diagonalizing Eq. \eqref{Sch NN}, $\mu_\downarrow$ and $\mu_\uparrow$ are the quasi-equilibrium chemical potential for, respectively, the spin down and spin up itinerant electrons \cite{Cywiski_07}. We have defined 
\begin{align*}
\mathcal{T}  =   &  2\pi e^{-\zb\frac{\zD}{2} }     \int_{\min(0, E_\downarrow -E_\uparrow  -\ze_m +\ze_{m+1}  ) }^\infty  \zr (E+E_\uparrow   +\ze_m -\ze_{m+1}  )     \zr (E+E_\uparrow)  \\&
\frac{1}{1+e^{\zb(E+E_\uparrow - \zm_\downarrow  +\ze_m -\ze_{m+1} ) }}
\frac{1 }{1+e^{-\zb(E+E_\uparrow-\mu_\uparrow) }}     \dif E \;,
\end{align*}  
where $\zr$ is the density the delocalized states.
Equation \eqref{kolmog_1} is obtained under the hypothesis that the electron gas is in a quasi-equilibrium condition and a single isotropic band is considered. The details of the calculations are given in Appendix \ref{appendix_atom-free_coll}. Equation \eqref{kolmog_1} can be easily extended to include two or more electron bands. Direct consequence of the balance equation is $W^{m}_0=W^{m+1}_-+W^{m-1}_+$. This ensures that the evolution equation describes a Markov chain and Eq. \eqref{kolmog_1} a Kolmogorov equation. From a physical point of view, Eq.  \eqref{kolmog_1} is obtained by assuming that the electron momentum and spin after the collision are uncorrelated with respected to their values before the collision. The choice of the out-coming momentums and spins is constrained only by the conservation of the total angular momentum and energy. Equation \eqref{kolmog_1} can be solved by applying two different techniques: the deterministic approach or the MC approach \cite{Muscato_16,Coco_16}. In the deterministic case Eq. \eqref{kolmog_1} is solved directly and the evolution of the density $n_m$ is obtained by applying some numerical discretization scheme \cite{Mascali_17,morandi_PRB_11}. To our purposes, the MC procedure is particularly suited. Compared to the deterministic solvers, the MC method provides some additional flexibility which results to be very useful in order to treat at the same time the collisions itinerant-localized electrons and the coherent evolution of the atomic states.    
In the MC framework, the time continuous density $n(t)$ is substituted by  a stochastic continuous-time process. The evolution of $n(t)$ is obtained by sampling the paths of the family of random variables $(X_t)_{t\in \mathbb{R}^+}$. The probability density of the continuous-time stochastic process  is directly related to $n(t)$. The link between the process $(X_t)_{t\in \mathbb{R}^+}$ and the  occupation probability $n(t)$ is given by the following result: we show that $\mathbb{E} (X_t)$, the expectation of $(X_t)_{t\in \mathbb{R}^+}$, is equal to the atomic magnetization obtained by solving the deterministic   Eq. \eqref{kolmog_1}.  
It is convenient to write the solution of Eq. \eqref{kolmog_1} in terms of the propagator $P(t)$
\begin{align}
 n_m (t) = P_{m,m'} (t)  n_{m'} (0)\; .\label{propag_n}
\end{align} 
The evolution equation for $P$ is
\begin{align}
 \dpt{P}{t}   =  Q P \label{evol P} \;,
\end{align} 
where infinitesimal generator $Q$ is defined in Eq. \eqref{kolmog_1} and the initial condition is  $P_{n,m} (0)=\zd_{n,m}$. We interpret the propagator $P$ as the transition semigroup of the continuous-time Markov process $(X_t)_{t\in \mathbb{R}^+}$. The family of random variables $(X_t)_{t\in \mathbb{R}^+}$ is a Poisson process assuming the discrete values $\{-S,-S+1,\ldots, S\}$. 
We associate $(X_t)_{t\in \mathbb{R}^+}$ to the evolution equation \eqref{kolmog_1}, by imposing  that the probability that $(X_t)_{t\in \mathbb{R}^+}$ moves from the value $m'$  to the value $m$ after a time delay $s$ is 	\cite{Privault_book} 
 \begin{align}
 \mathbb{P} \left( X_{t+s} = m |X_{t} = m' \right)  =P_{m,m'}(s) \;. \label{Prob X}
 \end{align} 
Since from Eq. \eqref{kolmog_1}, $\dpt{n_m}{t}$ depends only by $n_{m\pm1}$, $(X_t)_{t\in \mathbb{R}^+}$ jumps through consecutive values of the spin.
We denote by $(T_n)_{n\geq 1}$ the increasing family of jump times of $(X_t)_{t\in \mathbb{R}^+}$ 
\begin{align*}
T_1  = & \mathrm{inf} \left\{ t>0 |X_{t} \neq X_{0} \right\} \\
T_{n+1}  =& \mathrm{inf} \left\{ t>T_n |X_{t} \neq X_{T_n} \right\}\;. 
\end{align*} 
Let 
\begin{align*}
\tau_n  = & T_{n+1} -{T_n}  
\end{align*} 
denote the time between two consecutive jumps. 
By construction, the probability distribution $P$ associated to the jump times $T_n$ is solution of Eq. \eqref{evol P} which is of Kolmogorov-Markov type. As a consequence, a well known property of the Markov process guarantees that the sequence $(\tau_n)_{n\geq 1}$ is made of independent random variables  exponentially distributed. We have
\begin{align}
 \mathbb{P} \left( \tau_n > t | X_{T_{n-1}} =m\right) = e^{- W_0^m t}\;. \label{prob_exp_W}
\end{align} 
The total collision frequency $W_0^m$ governs the exponential decay of the probability. We verify the compatibility between the continuous description of the evolution   probability provided by Eq. \eqref{kolmog_1} and the statistic approach based on the stochastic  process $(X_t)_{t\in \mathbb{R}^+}$. 
We assume that at the time $t_0$ only the state $m'$ is populated 
 \begin{align*}
 n_{m}(t_0)&= \zd_{m,m'} \;.
 \end{align*}
Coherently, we require the initial condition $X_{t_0} = m'$. We have
\begin{align*}
\mathbb{E} \left( X_{t} = m |X_{t_0} = m' \right)&= \sum_m m \mathbb{P} \left( X_{t} = m |X_{t_0} = m' \right)= \sum_m m P_{m,m'}(t-t_0) 
\end{align*} 
where   we have used Eq. \eqref{Prob X}.
From  Eq. \eqref{propag_n} and Eq. \eqref{magn spin dens} we have 
\begin{align*}
S_z (t)  &=\sum_m mn_{m}(t)=\sum_m m P_{m,m'} (t-t_0) \;.
\end{align*}
We obtain that  expectation of $(X_t)_{t\in \mathbb{R}^+}$ is equal to the quantum mechanical expectation value of the atomic spin obtained by Eq. \eqref{kolmog_1}
\begin{align*}
\mathbb{E} \left( X_{t_0+t} = m |X_{t} = m' \right)&=S_z (t) \;.
\end{align*} 
From Eq. \eqref{kolmog_1} and Eq. \eqref{prob_exp_W} we obtain the collision frequencies 
\begin{align*}
\tau^{-1}_{l-i}(m)  \equiv 
&  W_0^m  = \mathcal{T} \left(w_{-}^m   e^{-\zb\frac{\zD}{2} }   + w_{+}^{m}    e^{ \zb\frac{\zD}{2} }    \right) \\=&   \frac{\zg^2}{\hbar} \mathcal{T} \left[ \left(S(S+1)-m^2 \right)  \left(  e^{\zb\frac{\zD}{2} } +  e^{-\zb\frac{\zD}{2}} \right)      + m   \left(  e^{\zb\frac{\zD}{2} } -  e^{-\zb\frac{\zD}{2} }\right)     \right]  \;.
\end{align*} 
For the sake of clarity, we write explicitly the infinitesimal generator defined in Eq. \eqref{kolmog_1} 
\begin{align*}
Q= \mathcal{T}\left(\begin{array}{ccccccccc}
W^{S}_0   &  W^{S}_-    &0    &0&0&0& \ldots \\
\vdots \\
\ldots  &      W^{m+1}_+  &W^{m+1}_0   &  W^{m+1}_-    &0    &0& \ldots \\
\ldots &0   &   W^m_+  &W^m_0    &  W^m_- &0& \ldots \\
\ldots & 0 & 0 &    W^{m-1}_+  &W^{m-1}_0   &  W^{m-1}_-  &   \ldots \\
\end{array}\right) \:.
\end{align*} 
We define
\begin{align*}
P_{(m,\uparrow )\rightarrow(m+1,\downarrow )} \equiv & \lim_{h \downarrow 0} \mathbb{P} \left(X_{t+h}= m +1  | X_{t}= m \textrm{ and } X_{t+h}-X_{t}\neq 0\right)\\
P_{(m, \downarrow)\rightarrow(m-1,\uparrow )} \equiv & \lim_{h \downarrow 0} \mathbb{P} \left(X_{t+h}= m -1  | X_{t}= m \textrm{ and } X_{t+h}-X_{t}\neq 0\right)\;,
\end{align*}
which represent the probabilities to have a transition from the initial state $m$ to one of the two possible states $m\pm1$, assuming that the collision takes place at the time $t$. We have  
\begin{align}
P_{(m,\uparrow )\rightarrow(m+1,\downarrow )} =  & \frac{W_{-}^{m+1}       }{ W_{-}^{m+1}+ W_{+}^{m-1}} = \frac{  1  }{ 1         +\frac{w_{-}^{m-1}}{w_{+}^{m+1}    }  e^{ \zb \zD }   } = \frac{  1  }{ 1         +\frac{ 1+ \frac{m}{S(S+1)-m^2} }{1- \frac{m}{S(S+1)-m^2}    }  e^{ \zb \zD }   }  \quad ; m\neq S \label{P m_mp}\\
P_{(m,\downarrow )\rightarrow(m-1,\uparrow )} = &\frac{  1  }{ 1         +\frac{ 1- \frac{m}{S(S+1)-m^2} }{1+ \frac{m}{S(S+1)-m^2}    }  e^{- \zb \zD }   } \quad ; m\neq -S\label{P m_mm}\\
P_{(-S,\uparrow )\rightarrow(-S+1,\downarrow )} = &  P_{(S,\downarrow)\rightarrow(S-1,\uparrow  )} = 1\label{P m_S}  \;.
\end{align}

 \section{Two-body interaction: first order approximation}\label{sec_two-b int}
 In the previous sections we have introduced the relevant dynamic equations for the atom-atom spin interaction at the molecular field level (Sec.  \ref{sec_NN_int}) and for the scattering among localized and itinerant electrons (Sec. \ref{sec_atom_free_int}). There are important physical processes that are not included in the previous cases. In particular, we are interested  to improve the description of the   exchange interaction and  spin transfer  by including some effects beyond the  molecular field approximation. The spin exchange mechanism is a  two-particle process.  The Hilbert space is the tensorial product of all the single particle electron spin spaces and the relevant  mathematical description has been developed in the framework of the many body theory of dynamical systems. A complete many body treatment of the problem lead to a mathematical formulation which is usually beyond to the numerical possibilities.
 Different methods have been developed in order to simplify the problem. Typically, the many body solution is projected on the single particle atomic state space. In this way, the atom ensemble is described by a relatively small Hilbert space. 
 The standard way to achieve this decoupling is the application of the Fermi Golden Rule (FGR). The FGR provides the first order probability to have a transition from a initially occupied state to another available state after a time which is typically large compared to the interaction time. The relevant expansion parameter is the interaction strength. This procedure is particularly suited if the final states form a continuous, as for example  the energy band structure in a solid. In fact, the FGR is at the basis of the derivation of Eq. \eqref{kolmog_1}.
 A second point is important for our discussion, we remind that the application of the FGR assumes implicitly that the typical collision time is short.
 From a mathematical point of view, the scattering probability is calculated by neglecting the  time scale during which the collision takes place. It is assumed that the initial state is populated at the time $-\infty$ and the final state is observed at the time $+\infty$. In this limit, the time-energy  uncertainly disappears and we recover the conservation of the energy during the collision expressed in the form of a Dirac delta function. In order to make the description consistent, it is thus necessary to have a population of particles distributed on a continuous of states, over which the delta function can be calculated. 
 In the same time, the application of the FGR ensures that the final process be Markovian. In fact, having taken the limit from $-\infty$ to $+\infty$ for the collision time interval, all the memory effects are expected to disappear. 
 
 In this section, we modify the standard FGR procedure in order to describe the direct atom-atom interaction beyond the molecular field approximation. The standard perturbation approach does not apply straightforwardly and some modifications to the standard FGR procedure are necessary. The reason is that the atomic spin states are usually a small number of discrete states and cannot be approximated by a continuous. Coherently with the phenomenas that we have investigated so far, the final result will be interpreted in the probabilistic way, treated as an additional stochastic process and simulated by the MC approach. 
 %

 We derive the transition probability for the two-particle system in the presence of exchange interaction at the first order on the perturbation expansion.
 We consider two atoms in interaction and assume that the atoms are described by the following single particle wave functions, respectively  $\zy^{\tiny\textcircled{a}}$ and $\zy^{\tiny\textcircled{b}}$. We fix the notation. We define by $\chi^{\tiny\textcircled{a}}_m$  and  $\chi^{\tiny\textcircled{b}}_j$, respectively the basis elements for the system $\textcircled{a}$ and $\textcircled{b}$. In Dirac notation 
 \begin{align*}
 \left|  \zy^{\tiny\textcircled{a}}(t) \right\rangle &=\sum_{m=1}^{N^{\tiny\textcircled{a}}} \left|\chi^{\tiny\textcircled{a}}_m \right\rangle \zy^{\tiny\textcircled{a}}_m (t) \;. 
 \end{align*}
 The two-particle wave function is defined by $
 \Psi =  \zy^{\tiny\textcircled{a}} \otimes \zy^{\tiny\textcircled{b}}$ and componentwise $
 \Psi_{mj} =  \zy^{\tiny\textcircled{a}}_m \zy^{\tiny\textcircled{b}}_j $. 
 The basis elements for the two-particle system are $\left|\chi^{\tiny\textcircled{a}}_m ,\chi^{\tiny\textcircled{b}}_j \right\rangle^{\tiny\textcircled{a}\otimes\tiny\textcircled{b}}$. To compact notations it is convenient to introduce a collective index $r=\left\{m,j\right\}$ and
  \begin{align*}
  \left|  \Psi (t) \right\rangle &=\sum_{r=1}^{N^{\tiny\textcircled{a}\otimes\tiny\textcircled{b}}} \left|\chi_r \right\rangle^{\tiny\textcircled{a}\otimes\tiny\textcircled{b}} \Psi_r (t) \;.
  \end{align*}
 We denote by $\mathcal{H}^{l-l}$ the two-particle Hamiltonian that describes the  interaction. The many-body quantum dynamics is usually extremely hard to solve and the complexity increases exponentially with the number of particles. We limit ourselves to consider the first order transition induced by $\mathcal{H}^{l-l}$ in the two-particle case.
 The exchange Hamiltonian is defined as follows
 \begin{align}
 \mathcal{H}_{m_1,j_1;m_2,j_2}^{l-l} = \zg \mathbf{J}^{\tiny\textcircled{a}}_{m_1,j_1} \cdot   \mathbf{J}^{\tiny\textcircled{b}}_{m_2,j_2} \label{H2p_exch}\;, 
 \end{align}
 where the dot denotes the scalar product, $\mathbf{J}$  the spin matrices and $\zg$ the exchange coupling constant.  Our results can be easily extended to more general two-particle interactions.
 %
 %
 %
 Fixing the $z-$axis as quantization axis,  Eq. \eqref{H2p_exch} becomes 
 \begin{align*}
 \mathcal{H}_{m_1,j_1;m_2,j_2}^{l-l} = &\zg \; \zd_{j_1,m_1}\zd_{j_2,m_2} m_1 m_2  +\frac{\zg}{2}\left([ {J}_-^{\tiny\textcircled{a}}]_{m_1,j_1}  [ {J}^{\tiny\textcircled{b}}_+]_{m_2,j_2}   +   [ {J}^{\tiny\textcircled{a}}_+]_{m_1,j_1}  [ {J}_-^{\tiny\textcircled{b}}]_{m_2,j_2}\right)\;,
 \end{align*}
 where $S$ denotes the total atomic spin and we have defined the operators $ {J}_+= {J}_x+i  {J}_y$ and $ {J}_-= {J}_x-i  {J}_y$ 
 \begin{align*}
 [ {J}_\pm]_{j,j'} &=  \zd_{j,j'\pm1}  \sqrt{ S ( S+1)-jj'}  
 \;.
 \end{align*} 
 For $t=0$ we assume that only one component of the two-particle wave function is populated. 
  \begin{align*}
  \left|  \Psi (t_0) \right\rangle &=\left|\chi_{r_0} \right\rangle^{\tiny\textcircled{a}\otimes\tiny\textcircled{b}} \;. 
  \end{align*}
 First order perturbation theory gives
 \begin{align}
   \left|  \Psi_r (t) \right|^2 =& \frac{4}{ \left|  E_r-E_{r_0} \right|^2 } \left|   \mathcal{H}^{l-l}_{r_0,r} \right|^2 \sin^2 \left(\frac{(E_r-E_{r_0})t}{2\hbar}\right) ^2\simeq \frac{\left|   \mathcal{H}^{l-l}_{r_0,r} \right|^2 }{\hbar^2   }  t^2 \label{FGR_Sch}\;,
\end{align}
 where
 \begin{align*}
 \mathcal{H}^{l-l}_{r_0,r} =&  \left\langle\chi_{r_0}  \left|\mathcal{H}^{l-l} \right|\chi_r \right\rangle^{\tiny\textcircled{a}\otimes\tiny\textcircled{b}} \;.
 \end{align*}
 This result can be expressed by the density matrix point-of-view. The two-particle density matrix is
  \begin{align*}
  \zr^{\tiny\textcircled{a}\otimes\tiny\textcircled{b}}  =& \sum_{r,s}\left|\chi_s \right\rangle\left\langle\chi_r \right| ^{\tiny\textcircled{a}\otimes\tiny\textcircled{b}}    \zr_{r,s} (t)   \;,
  \end{align*}
and $
\zr^{\tiny\textcircled{a}\otimes\tiny\textcircled{b}}(t_0)  = \left|\chi_{r_0} \right\rangle\left\langle\chi_{r_0} \right| ^{\tiny\textcircled{a}\otimes\tiny\textcircled{b}} $. Equation \eqref{FGR_Sch} gives the first order diagonal transitions
  \begin{align*}
  n_r(t) \equiv\zr_{r,r}(t) \simeq& \frac{\left|   \mathcal{H}^{l-l}_{r_0,r} \right|^2 }{\hbar^2   }  t^2\;.
  \end{align*}
Up to the first order in time, this result can be conveniently descried in term of a Markov process if we rescale the time $\overline{t}=t^2$ and
  \begin{align*}
\overline{n}_r  \left(\overline{t}\right) \equiv    n_r  \left(\sqrt{\overline{t}}\right) =& \frac{\left|   \mathcal{H}^{l-l}_{r_0,s} \right|^2 }{\hbar^2   }  \overline{t}\;.
  \end{align*}
  Up to the first order in the time variable $\overline{t}$ and with initial conditions $n^a_m = \zd_{m,m_1}$, $n^b_m = \zd_{m,m_2}$, the evolution equations for the densities are (analogous for the state $b$)
  \begin{align}
  \dpt{\overline{n}_{m_1}^a}{\overline{t}} =& -\frac{1}{\hbar^2}\left( \left|\mathcal{H}_{m_1,m_1+ 1;m_2,m_2- 1}^{l-l}\right|^2+\left|\mathcal{H}_{m_1,m_1- 1;m_2,m_2+ 1}^{l-l} \right|^2 \right) n_{m_1}^a \label{Mark_cohe_1} \\ 
  \dpt{\overline{n}_{m_1+1}^a}{\overline{t}} =&\frac{1}{\hbar^2}\left|\mathcal{H}_{m_1,m_1+ 1;m_2,m_2- 1}^{l-l}\right|^2 n_{m_1}^a\label{Mark_cohe_2}\\
  \dpt{\overline{n}_{m_1-1}^a}{\overline{t}} =&\frac{1}{\hbar^2}\left|\mathcal{H}_{m_1,m_1- 1;m_2,m_2+ 1}^{l-l} \right|^2  n_{m_1}^a\label{Mark_cohe_3}\;. 
  \end{align}
Equations  \eqref{Mark_cohe_1}-\eqref{Mark_cohe_3} are a Kolmogorov system. Me apply the same procedure described in Sec. \ref{sec_atom_free_int} and we obtain an approximate solution of the system in terms of a continuous-time stochastic process.
For each initial value of the spin of the atoms $a $ and $b$, denoted respectively by $m$ and $m'$ we have two transitions $(m,m')\rightarrow(m\pm1,m'\mp1)$.
The collision times related to such transitions are (we remind that we are using the scaled time variable $\overline{t} = t^2$) 
\begin{align*}
\frac{1}{\overline{\tau}_{l-l}(m,m')} =& \frac{1}{\hbar^2}\left( \left|\mathcal{H}_{m,m+ 1;m',m'- 1}^{l-l}\right|^2+\left|\mathcal{H}_{m,m- 1;m',m'+ 1}^{l-l} \right|^2 \right)\\
=&\frac{\zg^2}{2\hbar^2} \left[\left(S^a ( S^a+1) -m^2 \right) \left(S^b ( S^b+1) -{m'}^2 \right) - m m'\right]  \;.
\end{align*}
We have used
\begin{align*}
\left|  \mathcal{H}_{m,m\pm 1;m',m'\mp 1}^{l-l}\right|^2 = 
   &\frac{\zg^2}{4} \left(S^a ( S^a+1) -m (m\pm 1)\right) \left(S^b ( S^b+1) -m' (m'\mp 1)\right)   \;. 
   \end{align*}
We indicate by  $\overline{\tau}_k =\overline{T}_{k+1}- \overline{T}_k $ the time spend in the $i-$th state between two consecutive collisions. Due to Eqs. \eqref{Mark_cohe_1}-\eqref{Mark_cohe_3}, the sequence $(\overline{\tau}_i)_{i\geq1}$ is made of independent random variables having the same exponential distribution. Each collision is described in the original time variable $t$ by the collision times $T_k$, and $\tau_k =T_{k+1}- T_k $. We have
\begin{align}
\mathbb{P} \left(  \tau_k > t^2 | X^a_{T_{k}}=m,X^b_{T_{k}}=m' \right)  = \mathbb{P} \left( \overline{\tau}_k > \overline{t} | \overline{X^a}_{\overline{T}_{k}}=m,\overline{X^a}_{\overline{T}_{k}}=m' \right) = e^{-  \frac{t^2}{\overline{\tau}_{l-l}}}\;.  
\end{align} 
We have denoted by $(X^a_t)_{t\in \mathbb{R}^+}$ and $(X^b_t)_{t\in \mathbb{R}^+}$ the time-continuous stochastic processes associated respectively to the state $a$ and $b$.
We are also interested to the probability that one collision is produced in the interval $n\Delta_t < t< (n+1)\Delta_t$, where $\Delta_t$ denotes the time step related to the variable $t$. 
\begin{align*}
&\mathbb{P} \left( \tau_k >  (n+1)\Delta_t  \big| \tau_k > n\Delta_t  \textrm{ and } X^a_{T_{k}}=m,X^b_{T_{k}}=m'\right)\\=&\mathbb{P} \left(\overline{\tau}_k >  (n+1)^2\Delta_t^2  \big| \overline{\tau}_k > n^2\Delta_t^2 \textrm{ and } \overline{X^a}_{\overline{T}_{k}}=m,  \overline{X^b}_{\overline{T}_{k}}=m'   \right)\\
=&\frac{\mathbb{P} \left(\overline{\tau}_k >  (n+1)^2\Delta_t^2  \big|  \overline{X^a}_{\overline{T}_{k}}=m,  \overline{X^b}_{\overline{T}_{k}}=m' \right)}{\mathbb{P} \left(\overline{\tau}_k > n^2\Delta_t^2    \big|   \overline{X^a}_{\overline{T}_{k}}=m,  \overline{X^b}_{\overline{T}_{k}}=m' \right)}=e^{-\frac{ (n+1)^2\Delta_t^2   -n^2\Delta_t^2}{\overline{\tau}_{l-l}}}=e^{-\frac{ \Delta_t^2 }{\widetilde{\tau} }}\;, 
\end{align*} 
where we have defined $\widetilde{\tau} =\frac{\overline{\tau}_{l-l}}{2n+1 }$. 
Finally, the probability transitions are
\begin{align}
\lim_{h\downarrow 0} \mathbb{P} ( X_{t+h}^a = m \pm 1  | X^a_{t} = m\textrm{ ,  }X^b_{t} = m' \textrm{ and  } X_{t+h} - X_{t} \neq 0  )=
\frac{\overline{\tau}_{l-l}}{\hbar^2} \left|\mathcal{H}_{m,m\pm 1;m',m'\mp 1}^{l-l}\right|^2  \;. \label{P_coh} 
\end{align}

The case of spin 1/2 particle is particularly relevant. In this case the spin-spin Hamiltonian takes the form
\begin{align*}
\mathcal{H}^{l-l} =& \frac{\zg}{2}\left(  \zs_+ \otimes    \zs_- + \zs_- \otimes    \zs_+\right) \;, 
\end{align*}
where $ \zs_+= \zs_x+i  \zs_y$, $\zs_-= \zs_x-i \zs_y$ and $\zs_i$, $i=x,y,z$ denote the  Pauli matrices.  In the two-particle basis states $\left|\uparrow ,\uparrow  \right\rangle $, $\left|\uparrow ,\downarrow  \right\rangle $, $\left|\downarrow ,\uparrow  \right\rangle $, $\left|\downarrow ,\downarrow  \right\rangle $, the Hamiltonian becomes
\begin{align*}
\mathcal{H}^{l-l} =& \frac{\zg}{2} \left( \begin{array}{cccc}
0&0&0&0\\
0&0&1&0\\
0&1&0&0\\
0&0&0&0\\
\end{array} \right)  \;. 
\end{align*}
The collision time is for the transitions $\left|\uparrow ,\downarrow  \right\rangle  \rightarrow  \left|\downarrow ,\uparrow  \right\rangle $ and $  \left|\downarrow ,\uparrow  \right\rangle\rightarrow \left|\uparrow ,\downarrow  \right\rangle  $ is
\begin{align*}
\frac{1}{\overline{\tau}_{l-l}} =& \frac{\zg^2}{4\hbar^2}  \;,
\end{align*}
and the transition probabilities
 \begin{align*}
 {P}_{\left( \uparrow,\downarrow \right)\rightarrow \left( \downarrow,\uparrow \right)}=& {P}_{ \left( \downarrow,\uparrow \right)\rightarrow\left( \uparrow,\downarrow \right)} =1\;.
 \end{align*}

 \section{Implementation of the Monte Carlo procedure }
 
In order to clarify the application of our MC model, we describe into details how we solve the dynamics of  magnetism in a solid. 
We have classified the atom interactions in two groups: the coherent and the collisional ones. The coherent dynamics describes the atomic spin-spin interaction at the molecular field level. The collisional dynamics describes the interaction between atoms and delocalized electrons (interaction type $l-i$) and the atomic spin-spin interaction beyond the molecular field approximation (interaction type $l-l$). We simulate the coherent dynamics by solving $N$ coupled Schr\"odinger equations. The coherent dynamics is interrupted by instantaneous scattering events estimated by random numbers. Once a collision takes place, we project the wave function of the atom affected by the collision on the basis states and we select a single state according to the  rules of the collapse of the wave function in quantum mechanics. 
We simulate the collision by generating random numbers. We select the post-collisional state according to the transition probabilities of all the accessible final states. After this, we close the loop by restarting the calculation of the coherent dynamics. We illustrate this procedure by adding few technical points. At the time $t_0$, for each atomic position $R_i$ we extract two random numbers $\ {x}_{l-i}$ and  $ {x}_{l-l}$ associated respectively to the  atom-itinerant electrons and atom-atom scattering. The random variables are in the interval $[0,1]$ with uniform probability. As discussed before, the  collisional times for the $l-i$ and $l-l$ processes are given by, respectively $\tau_{l-i}$ and $\overline{\tau}_{l-l}$.  We estimate that the atom at the position $R_i$  will experience a collision  at the time $t^*=t_0 + \min (t_{l-i},t_{l-l} ) $, where $t_{l-l} = \sqrt{-\overline{\tau}_{l-l} \ln( {x}_{l-l})}$ and $t_{l-i}=-\tau_{l-i} \ln( {x}_{l-i}) $. During the interval $[t_0,t^*)$ we solve Eq. \eqref{Sch NN} numerically and we calculate $\zy(t^*)$. The initial state for the collision is obtained by the discrete random variable $x$ which is in the set $\{-S,\ldots, S \}$ with probability $|\zy_m(t^*)|^2$.  In the case $t_{l-i} < t_{l-l}$, the collision is of type $l-i$. In this case we have two possible transitions, respectively $(m,\uparrow)\rightarrow(m+1,\downarrow)$ and $(m,\downarrow)\rightarrow(m-1,\uparrow)$ with probabilities given by Eqs. \eqref{P m_mp}-\eqref{P m_S}.

In the case the dominant collision is of type $l-l$, we evaluate the spin of the neighbor atom with which the spin exchange process is done. We proceed the same way as we have done for the $i-$th atom. If we denote by $m'$ the spin of the neighbor atom, we have two possible transitions $(m,m')\rightarrow(m\pm1,m'\mp1)$ with probabilities given by  Eq.  \eqref{P_coh}. We select one of these two possibility by generating a random number and we update the spins states of both the atoms.

 \section{Numerical results}
 
We illustrate our approach by calculating the static magnetization of a homogeneous sample  constituted by a single specie of atoms arranged in a cubit lattice. We perform our calculations for Cobalt (fcc). Each atom is represented by a single wave function $\zy^i (t)\in \mathbb{C}^{2S+1}$ where the total atomic spin $S=1$. 

The main contribution to the magnetic order comes from the Heisenberg exchange interaction. For the sake of simplicity, we cutoff the exchange interaction up to the first six neighbors of each atom. Concerning the exchange constant we take $\zg=35.7 $ meV  which has been obtained by ab initio calculations in Ref. \cite{Pajda_01}.
%
  \begin{figure}[h]
  	\begin{center}
  		\includegraphics[width=0.49\columnwidth]{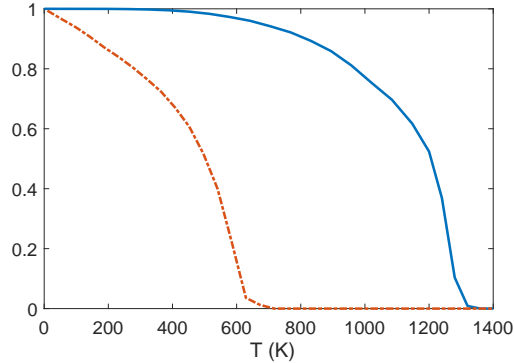}
  		\caption{Static magnetization of Co. Blue continuous curve: Quantum MC model. Red dashed curve: Classical LLG model.  \label{fig Co_stat_M}}
  	\end{center}
  \end{figure}
  \begin{figure}[h]
  	\begin{center}
  		\includegraphics[width=0.49\columnwidth]{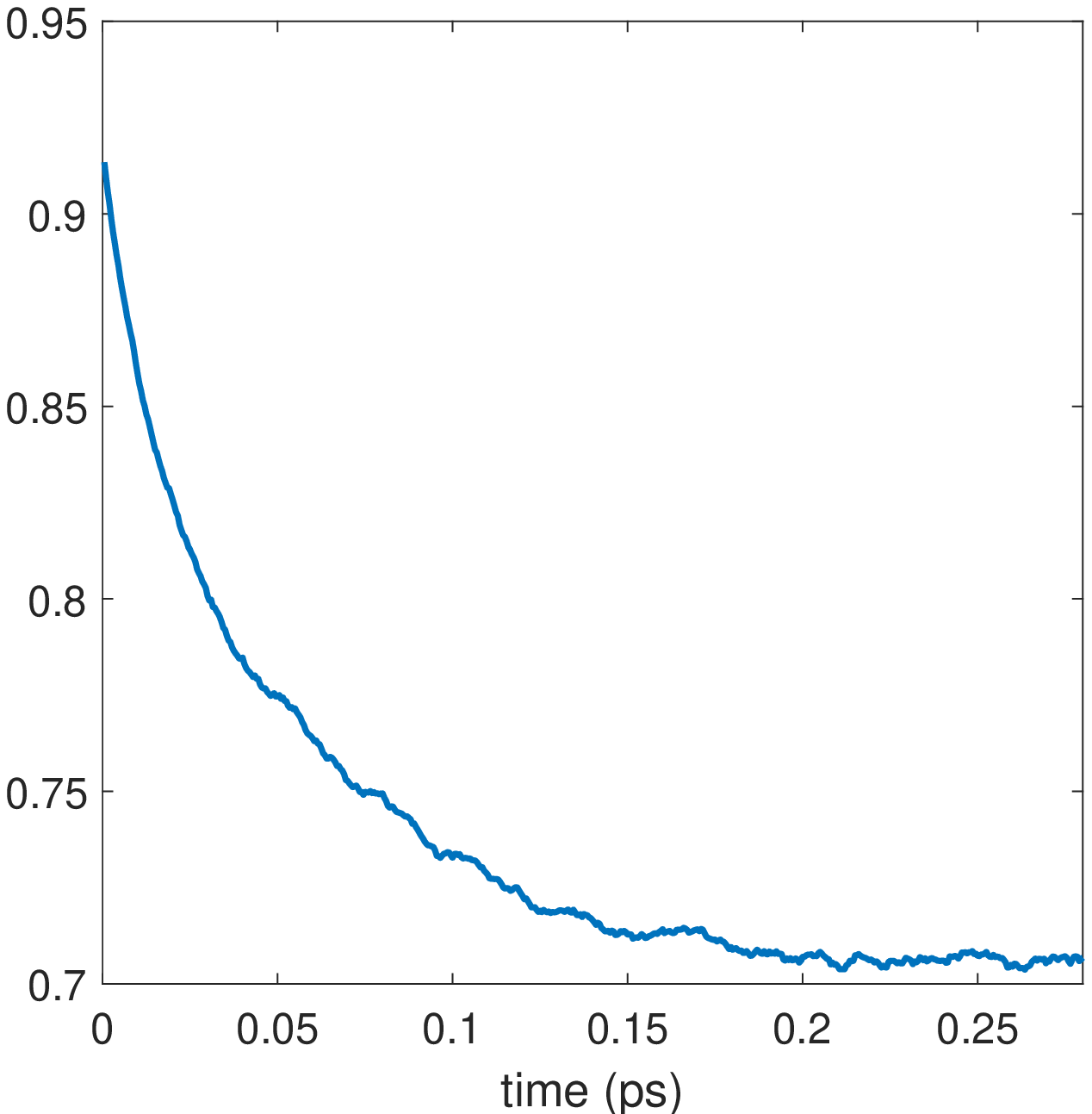}
  		\includegraphics[width=0.49\columnwidth]{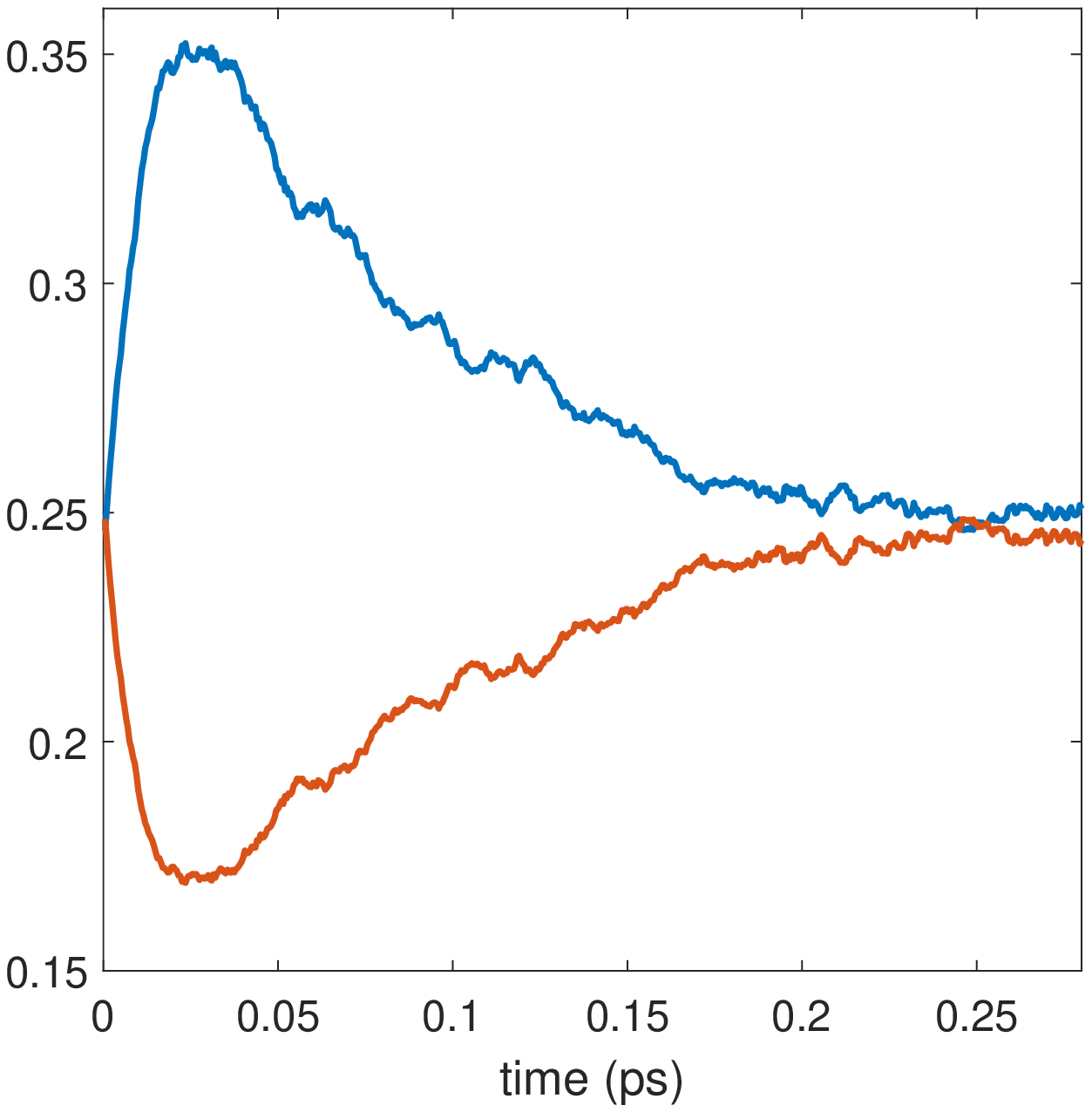}
  		\caption{Dynamics of the magnetization. Left panel: Mean magnetization of the sample. Right panel: Spin dependent chemical potentials,  $\mu_\uparrow$ (red curve) and $\mu_\downarrow$ (blue curve). \label{fig evol_M}}
  	\end{center}
  \end{figure}
We do not include an external magnetic field. 
Due to the axial symmetry of the interactions, the atomic spins are collinear and the local Hamiltonian is diagonal. The eigenvalues of the atomic Hamiltonian at the position $\mathbf{R}^i$ are
\begin{align*}
\ze^i_m  = -  m\;     \frac{ \zg }{  \hbar^2  }   \sum_{\langle   j\in \textrm{ NN}_i \rangle} S^j_z  \quad m = -S,\ldots , S.
\end{align*}
In figure \ref{fig Co_stat_M} (continuous blue curve) we depict the magnetization curve of Cobalt as a function of the temperature. Each stationary state is obtained by solving the time dependent system for a time interval sufficiently long. The number of atoms used in the simulations is around $7\times 10^5$. 
Our estimation of the Curie temperature $T_C$ agrees with the  experimental value  $T_C\simeq 1300$ K.
In order to appreciate the advantage of using a quantum model, in the same figure, we compare our results with the magnetization curve obtained by solving the classical LLG equation as described in Ref. \cite{Evans_14} (dashed red curve) with the same parameters. 

The magnetization dynamics is reproduced in figure 
\ref{fig evol_M}. The simulations are preformed as follows. As initial state we take the equilibrium state obtained by fixing the temperature of the system. We excite the system by rising instantaneously the temperature of the electron gas \cite{morandi_PRB_10,morandi_NJP_09,morandi_PRB_10b,morandi_PRB_11b}. This simulate the excitation of the system by an intense and short laser pulse. Typically, the laser pulses used for the study of the ultrafast magnetization dynamics in nanostructures have a width smaller than 100 fs. The main effect of such an excitation is to create a quasi-equilibrium population of hot electrons in the delocalized bands \cite{Beaurepaire_96,Bergeard_14}. The electrons located around the atoms are less affected by the excitation. In the left panel of figure \ref{fig evol_M} we depict the evolution of the magnetization of the sample after the excitation. In our simulation we have chosen as temperatures before and after excitation, respectively 800 K and 1150 K.  Our results agree with the experiments and reproduce the ultrafast demagnetization effect which has been observed in various nanometric systems \cite{Bigot_13,Bergeard_14,Boeglin_10,Radu_11,Vodungbo_12,Bigot_09,Stamm_07}. In the right panel we depict the evolution of the spin dependent chemical potential of the delocalized electrons $\mu_\uparrow$ (red curve) and $\mu_\downarrow$ (blue curve). As expected, at the equilibrium, the spin up and spin down chemical potentials coincide (see Eq.  \eqref{eq mu}). Our results show that the demagnetization dynamics is characterized by out-of-equilibrium distribution of itinerant electrons. 

%
%
%
%
 
\section{Conclusions}

We have presented a dynamical model that describes various types of atomic interactions inside a magnetic nanomaterial.  Our model is based on the assumption that the atoms form a net of interacting localized quantum particles that exchange energy and spin with gases of itinerant electrons. We divide the dynamics in two parts: coherent evolution and collisions. The coherent evolution is treated at the quantum level, the evolution of the atom wave function is evaluated by solving a set of nonlinear Schr\"odinger equations. The coherent dynamics of the atoms is interrupted by instantaneous collisions with itinerant electrons. We associate each collision event to the quantum collapse of the local atomic wave function. We reproduce the collision processes by a MC approach. In this contribution we have presented into details the algorithm used for the simulations. In particular, we have focused our discussion on the interplay and on the compatibility between the stochastic and the deterministic description of the particle dynamics. An innovative aspect of our research concerns the use of stochastic models to describe  the coherent two-particle interaction  beyond the local main field approximation. 
By reducing significantly  the computational effort required by the simulations, this procedure can be very useful to develop realistic models of systems containing many particles. 

Our model is potentially able to include many aspects of the microscopic atomic dynamics such as the distinction between spin and orbital angular momentum, the spin-orbit effect, the magnetic anisotropy, the orbital quenching due to the lattice potential. In Ref. \cite{Morandi_PRB17} we have compared our simulations  with the experiments presented in Ref.  \cite{Bergeard_14}, where  the ultrafast magnetization dynamics of Co$_x$Tb$_{1-x}$ has been investigated. 

Our stochastic-quantum model seems very promising for future applications. In fact, our approach is able to  reproduce two important aspects of the physics of open quantum systems containing many particles: the equilibrium configuration and the dynamics induced by external perturbations.

\appendix

\section{Collision between localized and itinerant electrons}\label{appendix_atom-free_coll}

The evolution equation for the localized spin particles in contact with a bath of itinerant electrons is given by 
\begin{align}
\left. \dpt{n_{m}}{t} \right|_{l-i} =
&    n_{m+1}        \mathcal{T}^{m+1,m}_{\downarrow ,\uparrow}    +   n_{m-1}     \mathcal{T}^{m-1,m}_{\uparrow,\downarrow}      -n_{m}      \left(\mathcal{T}^{m,m+1}_{\uparrow,\downarrow}  +\mathcal{T}^{m,m-1}_{\downarrow,\uparrow}  \right)   \;,\label{balance_nm}
\end{align} 
where 
\begin{align*}
\mathcal{T}^{m',m}_{\zs',\zs} =  &  2\pi    \int W_{m'\zs' \rightarrow m \zs }( \mathbf{k}' ,\mathbf{k}) \; 
[1-f_{\zs }(\mathbf{k})]  f_{\zs'}(\mathbf{k}')       \zd( E_{\zs } (\mathbf{k})-E_{\zs'} (\mathbf{k}')+\ze_m -\ze_{m'}  )\frac{\dif\mathbf{k}'}{(2\pi)^3}\frac{\dif\mathbf{k}}{(2\pi)^3} \; .
\end{align*}
We denote by $f_{\zs}(\mathbf{k})$ the electron distribution function and by $\mathbf{k}$ the electron wavenumber. The collision kernel is
\begin{align*}
W_{ m'\zs' \rightarrow m \zs  }( \mathbf{k}' ,\mathbf{k}) =&   w_{\pm}^m \zd_{m,m'\pm 1}\zd_{\zs, \zs'\mp 1} \;,
\end{align*}
with $ w_{\pm}^m  = \frac{\zg^2}{\hbar} \left( S(S+1)- m(m \mp 1)\right)$, $S$ is the total spin and $\zg$ is the exchange interaction strength. 
We assume isotropy of the bands. We write the energy of the spin up and spin down bands as 
$$E_{\zs } (\mathbf{k}) =E_{\zs } +\overline{E} (|\mathbf{k}|) \;,
$$
where $E_\zs$ is the band edge. 
We assume that $f$ is described by a quasi-equilibrium Fermi-Dirac distribution
\begin{align}
f_\zs(\mathbf{k}) = \left(1+e^{\zb (E_{\zs } (\mathbf{k})-\mu_\zs)  }\right)^{-1}\;, \label{FD dist}
\end{align}
where $\zm_\zs$ is the spin dependent chemical potential,  $\zb= \frac{1}{k_B T}$, $k_B$ is the Boltzmann's constant and $T$ is the gas temperature. 
%
With this assumption, we obtain
\begin{align*}
\mathcal{T}^{m',m}_{\zs',\zs} = \left\{
\begin{array}{ll}
\mathcal{C}^{m+1,m}_{\downarrow,\uparrow}&\mathrm{if }\; m'=m+1,\;\zs'=\downarrow,\;\zs=\uparrow\\
\mathcal{C}^{m-1,m}_{\uparrow,\downarrow} &\mathrm{if }\; m'=m-1,\;\zs'=\uparrow,\;\zs'=\downarrow\\
0 &\mathrm{otherwise}
\end{array}
\right. 
\end{align*}  
where 
\begin{align*}
\mathcal{C}^{m',m}_{\zs',\zs} =  &  2\pi  w_{m-m'}^m    \int_{0}^\infty  \zr (E'+E_{\zs'} )     \zr (E+E_\zs)  
\frac{1}{1+e^{\zb(E'+E_{\zs'} - \zm_{\zs'}) }}
\frac{e^{\zb(E+E_\zs-\mu_\zs) }}{1+e^{\zb(E+E_\zs-\mu_\zs) }}    \\&     \zd( E+E_\zs  -E'- E_{\zs'} +\ze_m -\ze_{m'}  )\dif E'\dif E\;. 
\end{align*}
Here, $\zr$ is the density of states and $\ze_m$ the energy of the $m-$th localized spin state. After few manipulations and  using $w_{+}^{m+1} =w_{-}^m$ we obtain
\begin{align*}
\mathcal{T}^{m ,m+1}_{\uparrow,\downarrow}  =  \mathcal{T}^{m+1,m}_{\downarrow,\uparrow}  e^{\zb(  - \ze_{m+1} + \ze_{m } -\mu_\downarrow+\mu_\uparrow ) }  \\
\mathcal{T}^{m-1,m}_{\uparrow,\downarrow}     =  \mathcal{T}^{m,m-1}_{\downarrow,\uparrow}  e^{\zb(  - \ze_{m } + \ze_{m-1 } -\mu_\downarrow+\mu_\uparrow ) }  \; .
\end{align*} 
The evolution equation becomes 
\begin{align}
\left. \dpt{n_{m}}{t} \right|_{l-i} =
&    n_{m+1}        \mathcal{T}^{m+1,m}_{\downarrow ,\uparrow}    +   n_{m-1}  \mathcal{T}^{m,m-1}_{\downarrow,\uparrow}  e^{\zb(  - \ze_{m } + \ze_{m-1 } -\mu_\downarrow+\mu_\uparrow ) }\nn    \\ &   -n_{m}      \left(\mathcal{T}^{m+1,m}_{\downarrow,\uparrow}  e^{\zb(  - \ze_{m+1} + \ze_{m } -\mu_\downarrow+\mu_\uparrow ) }    + \mathcal{T}^{m,m-1}_{\downarrow,\uparrow}   \right)   \;.\label{balance_nm2}
\end{align} 
In many cases the energy of the localized states is linear, $  \ze_{m }= C+ m B $ for some coefficients  $C$ and $B$. In this case, the coefficients appearing in Eq. \eqref{balance_nm2} can be expressed by a single integral
\begin{align*}
\mathcal{T}_{\downarrow,\uparrow} =   &  2\pi    \int_{\min(0, E_\downarrow -E_\uparrow  -\ze_m +\ze_{m+1}  ) }^\infty  \zr (E+E_\uparrow   +\ze_m -\ze_{m+1}  )     \zr (E+E_\uparrow)  \\&
\frac{1}{1+e^{\zb(E+E_\uparrow - \zm_\downarrow  +\ze_m -\ze_{m+1} ) }}
\frac{1 }{1+e^{-\zb(E+E_\uparrow-\mu_\uparrow) }}     \dif E\;.  
\end{align*}  
We obtain
\begin{align*}
\left. \dpt{n_{m}}{t} \right|_{l-i} =
&    \mathcal{T}_{\downarrow,\uparrow}  \left[ n_{m+1}    w_{-}^m      +   n_{m-1}  w_{+}^{m}        e^{\zb(  - \ze_{m } + \ze_{m-1 } -\mu_\downarrow+\mu_\uparrow ) }       -n_{m}      \left(w_{-}^m     e^{\zb(  - \ze_{m+1} + \ze_{m } -\mu_\downarrow+\mu_\uparrow ) }    + w_{+}^{m}        \right) \right]\;,
\end{align*} 
where we have used
\begin{align*}
\mathcal{T}^{m+1,m}_{\downarrow,\uparrow} =  &    w_{-}^m      \mathcal{T}_{\downarrow,\uparrow} \\    \mathcal{T}^{m,m-1}_{\downarrow,\uparrow} =  &    w_{-}^{m-1}     \mathcal{T}_{\downarrow,\uparrow}\;.
\end{align*} 
It is convenient to write the equation in  a more symmetric form
\begin{align}
\left. \dpt{n_{m}}{t} \right|_{l-i} =
&    \mathcal{T} \left[ e^{ \zb\frac{\zD}{2} } n_{m+1}    w_{-}^m      +   n_{m-1}  w_{+}^{m}        e^{-\zb\frac{\zD}{2} }       -n_{m}      \left(w_{-}^m   e^{-\zb\frac{\zD}{2} }   + w_{+}^{m}    e^{ \zb\frac{\zD}{2} }    \right) \right]\label{balance_nm3}\;, 
\end{align} 
where we have used $- \ze_{m+1} + \ze_{m }= - \ze_{m } + \ze_{m-1 }$ and we have defined $\zD=  \ze_{m+1} - \ze_{m } +\mu_\downarrow-\mu_\uparrow $ and  $\mathcal{T} = e^{-\zb\frac{\zD}{2} }  \mathcal{T}_{\downarrow,\uparrow}  $.
Equation \eqref{balance_nm3} admits the following  stationary equilibrium distribution
 \begin{align}
 n_m  =  &  \frac{ e^{-\zb \ze_{m}}}{\sum_m e^{-\zb \ze_{m}}}   \\
 \mu_\downarrow =&\mu_\uparrow  \;. \label{eq mu}
 \end{align}  

For completeness we describe evolution equation for the itinerant charges.
\begin{align*}
  \dpt{f_{s}(\mathbf{k})}{t}    =  &   \sum_{j,m,m'} n_{m} W_{ m'j \rightarrow m s  }  \int f_{ j}(\mathbf{k}') \left(1-f_{s}(\mathbf{k}) \right)\zd( E_{\zs } (\mathbf{k})-E_{\zs'} (\mathbf{k}')+\ze_m -\ze_{m'}  )\frac{\dif\mathbf{k}'}{(2\pi)^3}\\
-   & f_{s}(\mathbf{k}) \sum_{j,m,m'}n_{m'}  W_{  m s \rightarrow m'j } \int    \left(1-f_{j}(\mathbf{k}') \right)\zd( E_{\zs } (\mathbf{k})-E_{\zs'} (\mathbf{k}')+\ze_m -\ze_{m'}  )\frac{\dif\mathbf{k}'}{(2\pi)^3} \;.
\end{align*}

For our purposes, it is not necessary to solve this equation. In our model we assume that the electrons are in a quasi-equilibrium state, defined by Eq. \eqref{FD dist} where the   time dependent parameters are the electron temperature $T$ and the spin dependent  quasi-Fermi level $\zm_\zs$. 
All the informations that we need in order to evaluate the evolution of $T$ and $\zm_\zs$ can be obtained from our MC scheme by consistency arguments.   
Every time that  the spin of one atom is modified by a collision with an itinerant spin, due to the conservation of the total spin and energy, the same amount of spin and energy, with opposite sign, is modified in the electron gas. With this consideration, we can estimate the modification of the  energy and the number of spin flips which occur during a time interval $\Delta_t$. 

\end{document}